\documentclass[conference,a4paper]{IEEEtran}
\IEEEoverridecommandlockouts
\usepackage{cite}
\usepackage{multirow}
\usepackage{amsmath,amssymb,amsfonts}
\usepackage{amsthm}
\usepackage{fontawesome}
\usepackage{graphicx}
\usepackage{textcomp}
\usepackage{hyphenat}
\usepackage[table]{xcolor}
\usepackage{comment}
\usepackage{tikz}
\usepackage{pifont}
\usepackage[colorlinks]{hyperref}
\usepackage[english]{babel}
\usepackage{amsthm}
\usepackage{enumerate}
\newtheorem{theorem}{Theorem}[section]

\title{Theorem and Proofs}
\usepackage{empheq}
 \usepackage{amsmath}
\def\BibTeX{{\rm B\kern-.05em{\sc i\kern-.025em b}\kern-.08em
    T\kern-.1667em\lower.7ex\hbox{E}\kern-.125emX}}

\usepackage{mathtools}
\usepackage{placeins}

\usepackage{algorithm}
\usepackage{algpseudocode}

\usepackage{empheq}
\begin{document}

\title{Efficient and Encrypted Inference using Binarized Neural Networks within In-Memory Computing Architectures
\vspace{-4mm}} 

\author{
\IEEEauthorblockN{Gokulnath Rajendran\thanks{©2025 IEEE.  Personal use of this material is permitted.  Permission from IEEE must be obtained for all other uses, in any current or future media, including reprinting/republishing this material for advertising or promotional purposes, creating new collective works, for resale or redistribution to servers or lists, or reuse of any copyrighted component of this work in other works.}\IEEEauthorrefmark{1}, Suman Deb\IEEEauthorrefmark{2}, Anupam Chattopadhyay\IEEEauthorrefmark{1}}
\IEEEauthorblockA{\IEEEauthorrefmark{1}NTU, Singapore, \IEEEauthorrefmark{2}CREATE, Singapore}
gokulnat002@e.ntu.edu.sg, suman.deb@imperial.ac.uk, anupam@ntu.edu.sg \vspace{-6mm}}
\maketitle 

\begin{abstract}

Binarized Neural Networks (BNNs) are a class of deep neural networks designed to utilize minimal computational resources, which drives their popularity across various applications. Recent studies highlight the potential of mapping BNN model parameters onto emerging non-volatile memory technologies, specifically using crossbar architectures, resulting in improved inference performance compared to traditional CMOS implementations. However, the common practice of protecting model parameters from theft attacks by storing them in an encrypted format and decrypting them at runtime introduces significant computational overhead, thus undermining the core principles of in-memory computing, which aim to integrate computation and storage. This paper presents a robust strategy for protecting BNN model parameters, particularly within in-memory computing frameworks. Our method utilizes a secret key derived from a physical unclonable function to transform model parameters prior to storage in the crossbar. Subsequently, the inference operations are performed on the encrypted weights, achieving a very special case of Fully Homomorphic Encryption (FHE) with minimal runtime overhead. Our analysis reveals that inference conducted without the secret key results in drastically diminished performance, with accuracy falling below 15\%. These results validate the effectiveness of our protection strategy in securing BNNs within in-memory computing architectures while preserving computational efficiency.
\end{abstract}

\section{Introduction}

Deep neural networks (DNNs) are crucial for various applications and are valuable assets for businesses. These models are typically trained on proprietary and valuable data, and achieving high accuracy demands significant resources. Therefore, it is essential to protect DNNs as intellectual property. Trained models can be deployed on edge devices using key exchange mechanisms and encryption techniques, which help mitigate the risks of man-in-the-middle attacks. However, there remains a possibility of model parameter extraction attacks on these edge devices.

Binarized neural networks (BNNs) are a type of DNN where both weights and activations are limited to values of +1 and -1. This constraint reduces the computational resources needed during inference, leading to lower energy consumption. Recent research on accelerators has focused on in-memory computing architectures based on emerging device technologies, such as resistive RAM (RRAM), spin-transfer torque magnetic RAM (STT-MRAM), and spin-orbit torque magnetic RAM (SOT-MRAM). A recent study demonstrated that implementing BNNs in RRAM crossbar structures offers several advantages, including the ability for a single comparator to substitute both the energy-intensive analog-to-digital converter and the activation function logic \cite{kim2019memory}.

\textbf{BNN inference with RRAM crossbar:} Once the training phase of the BNN is complete, the resulting weight matrix \( W_{j,k}^{b} \) can be mapped to the RRAM crossbar for inference. An effective method involves using two RRAM devices arranged in a column format to represent a single weight value. For instance, a weight of +1 is achieved by setting the upper cell to a low resistance state (LRS) and the lower cell to a high resistance state (HRS). In contrast, a weight of -1 is represented by reversing this configuration. Input signals are encoded similarly, with +1 represented as (1,0) and -1 as (0,1). The activation function employed in the BNN is the sign function, which processes the output of the matrix-vector multiplication denoted as \( y_{k} \). This sign function is highly beneficial for implementing BNNs in RRAM crossbars because it can be realized with a comparator integrated into the crossbar, providing a more energy-efficient option compared to traditional ADC and activation logic. In BNNs, a batch normalization (BN) layer is often crucial. This layer can be adjusted as a threshold \( B_{k} \) within the activation function and, in the context of RRAM implementation, serve as a comparison input to the comparator.

\begin{figure}[!t]
\centerline{\includegraphics[width=0.5\textwidth]{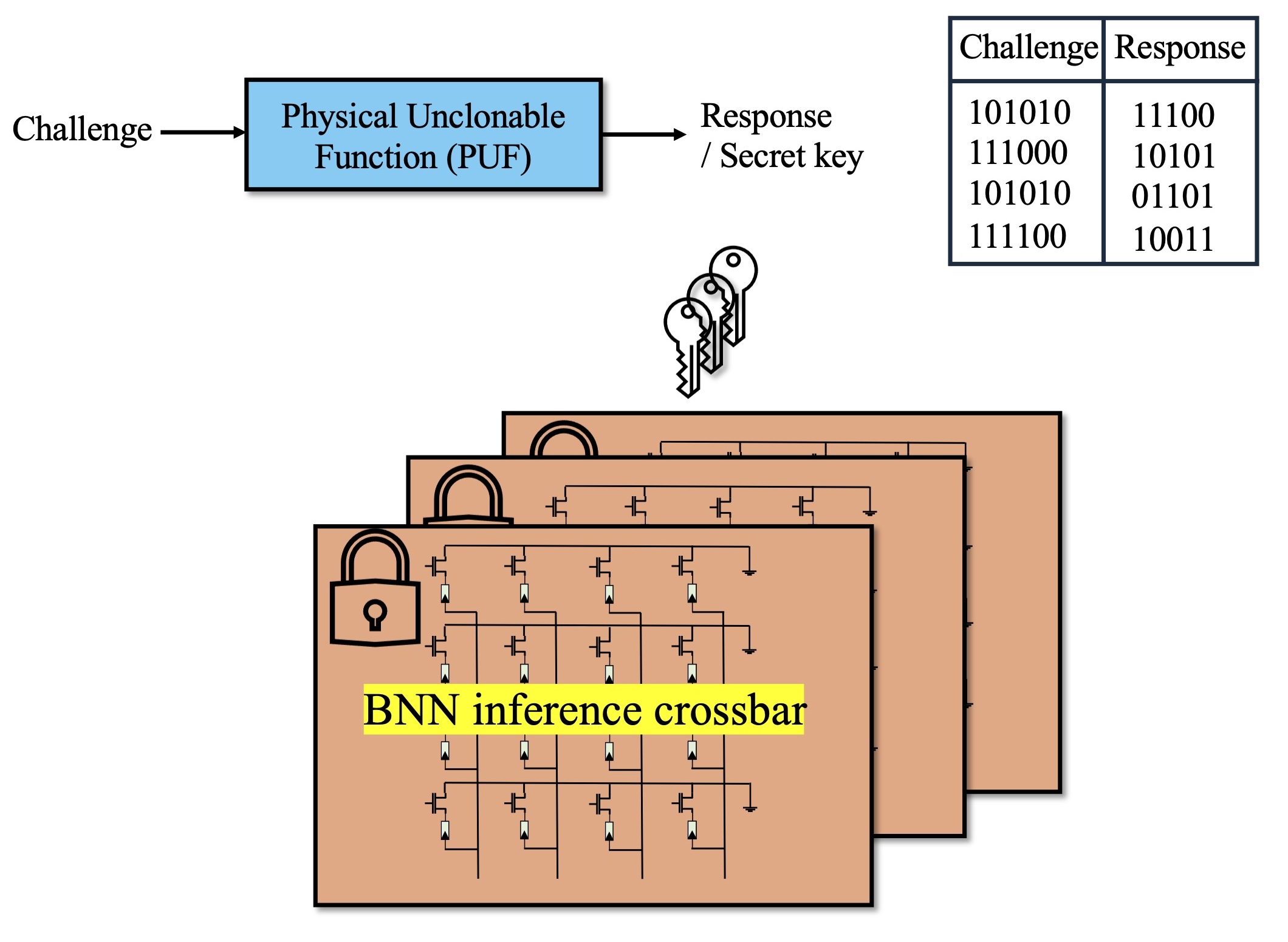}}
\caption{Overview of PUF Protected BNN Inference}
\label{protection}
\vspace{-6mm}
\end{figure}

\textbf{Related works:} Earlier studies have focused on protection techniques designed to prevent model theft attacks targeting multi-bit DNNs deployed on edge devices. The main idea is to transform the model parameters of the neural network with a secret key before mapping them onto crossbar structures. It is also essential to protect the extraction of the secret key used for such transformations. In our earlier work, we argued that secret keys generated through physical unclonable functions (PUFs) provide robust protection \cite{rajendran2024securing}. A PUF generates a response or secret key based on an input challenge applied during runtime, which means the key is not stored. This characteristic makes it more resilient against theft attacks, which typically aim to steal keys by accessing the state of non-volatile memory (NVM) devices. Therefore, it is preferred to use secret keys derived from PUFs to protect the model parameters, as detailed in Fig.\ref{protection}. Three techniques were previously discussed to protect multi-bit DNNs: shuffling \cite{galicia2023s3cure}, 1's complement \cite{zou2022enhancing}, and dummy weights \cite{wang2021low}. Both shuffling and 1's complement involve transforming the model parameters using keys, while dummy weights are used to hide the size of the weight matrix. We proposed two new techniques, namely inversion and swapping, in the context of BNN implementations in RRAM crossbars\cite{rajendran2024securing}. We also showed that dummy weights can be used not only to hide the size of the weight matrix but also as a method for weight transformation \cite{rajendran2024puf}.

The major contributions of this work are as follows:
\begin{itemize}
     
\item We formalize a method to protect any BNN inference implementation from model parameter theft attacks by using a secret key specifically derived from a PUF during runtime.

\item We introduce three new techniques for transforming model parameters based on our previous works. When these protected model parameters are used for inference without the secret key, the accuracy will significantly degrade.

\end{itemize}


\section{Encrypted BNN Inference}

\subsection{Attacker model}

The attacker can access the edge hardware equipped with an RRAM-based in-memory computing accelerator, where the trained model provided by the service provider is executed. By conducting reverse engineering, the attacker can uncover the topology of the neural network implementation within the RRAM hardware and retrieve its model parameters. However, the PUF and its responses in the edge device are consistently hidden and protected from the attacker at all times. As a result, the attacker cannot access the PUF or collect its responses. It is important to note that the secret keys are not stored as memory states in the PUF, unlike the model parameters. Instead, these keys are derived during runtime based on specific challenges.

\subsection{Theoretical Framework}

We present the following theorem that delineates strategies for safeguarding the parameters of the BNN model in any implementations. For proof of concept, we specifically reference RRAM technology. After training the BNN for its intended purpose, the resulting weight matrix \( W_{j,k}^b \) is directly loaded into the RRAM crossbar, which will then function as the inference platform. Protecting the confidentiality of the weight matrix \( W^b \) is critical due to its significance. We suggest transforming \( W^b \) into a new matrix denoted as \( W^{*b} \) utilizing a secret key generated through PUFs. This transformation ensures that, should \( W^{*b} \) be wrongfully acquired and utilized for inference without the corresponding secret key, it will be rendered ineffective. For the model to function correctly, it is crucial to employ the correct PUF key during the inference phase.

\begin{theorem}
Let a BNN layer be characterized by a weight matrix \( W \in \{\pm 1\}^{m \times n} \), a binarized input \( X \in \{\pm 1\}^{m} \), and an integer threshold \( B \in \mathbb{Z}^{n} \), which arises from Batch normalization integration. Then, the output \( Y \) of the unprotected layer is defined as:

\[
Y = \operatorname{sign}\left(W^\top X - B\right) \in \{\pm 1\}^{n}
\]


Let a PUF generate a hidden response \( R \in \{0,1\}^{d} \) that is accessible exclusively at runtime on the device and is never stored. We define the following reversible transformation which are parameterized by \( R \):

 \( \Gamma_R : \{\pm 1\}^{m \times n} \rightarrow \{\pm 1\}^{m \times n} \),
 \( \beta_R : \{\pm 1\}^{m} \rightarrow \{\pm 1\}^{m} \),
\( \delta_R : \mathbb{Z}^{n} \rightarrow \mathbb{Z}^{n} \) and
 \( \psi_R : \{\pm 1\}^{n} \rightarrow \{\pm 1\}^{n} \)

The stored (protected) model can then be represented as:

\[
(W^\star, B^\star) = \left(\Gamma_R(W), \delta_R(B)\right)
\]

During the inference phase, the device may apply \( \beta_R\) to inputs and \( \psi_R \) to the outputs generated by \( (W^\star, B^\star) \).

Then, for all \( X \in \{\pm 1\}^{m} \),
\begin{equation}
\psi_R\left(\operatorname{sign}\left(\Gamma_R(W)^\top \beta_R(X) - \delta_R(B)\right)\right) = \operatorname{sign}\left(W^\top X - B\right) \tag{†}
\end{equation}

We assert that distinct responses yield distinct protected models and recoveries when \( (W,B) \) is held constant. Specifically, for responses \( R \neq R' \),

\((\Gamma_R, \delta_R) \neq (\Gamma_{R'}, \delta_{R'})\) ,  \((\beta_R, \psi_R) \neq (\beta_{R'}, \psi_{R'})\).

\end{theorem}
\textbf{Proof:}
We present the proof for the case of column inversion, which extends to the general setting.

1. Consider a column \( k \) of a BNN defined by the equation 
\[
y_k = \sum_j W_{j,k} X_j, 
\]
where \( W_{j,k} \) are the weights and \( X_j \) are the inputs. The output bit is determined by 
\begin{equation}
y_k^{(b)} = \operatorname{sign}(y_k - B_k) \in \{-1, +1\},
\end{equation}
with the constraint that both \( y_k \) and \( B_k \) are even integers, ensuring the difference \( s = y_k - B_k \in 2\mathbb{Z} \). \\

2. Let \( R_k \in \{0, 1\} \) denote the hidden PUF bit for column \( k \). The mappings for weights and thresholds under transformation are given by:
\begin{equation}
W_{:,k}^{\star} = (-1)^{R_k} W_{:,k} 
\end{equation}
\begin{equation}
\quad B_k^{\star} = (1 - 2R_k) B_k + 2R_k.
\end{equation}
3. The protected inference output is expressed as:
\begin{equation}
y_k^{\text{out}} = (-1)^{R_k} \operatorname{sign}(y_k^{\star} - B_k^{\star})
\end{equation}
where 
\begin{equation}
y_k^{\star} = \sum_j W_{j,k}^{\star} X_j = (-1)^{R_k} y_k  
\end{equation}

From (5) and (3):
\begin{equation}
   y_k^{\star} - B_k^{\star} = (-1)^{R_k} y_k - \left((1-2R_k) B_k + 2R_k\right)
\end{equation}

   Thus, we simplify to:
\begin{equation}
   y_k^{\star} - B_k^{\star} = (-1)^{R_k} s - 2R_k.
\end{equation} \\ since \((-1)^{R_k}  = 1-2R_k \) \\

Thus (4) becomes
\begin{equation}
   y_k^{\text{out}} = (-1)^{R_k} \operatorname{sign}((-1)^{R_k} s - 2R_k).
\end{equation}

4. Case analysis:
\begin{enumerate}[i]
\item If \( R_k = 0 \):  \( y_k^{\text{out}} = \operatorname{sign}(s) \). 
\item If \( R_k = 1 \): \( y_k^{\text{out}} = -\operatorname{sign}(-s - 2) \). Since \( s \in 2{Z} \), either \( s \geq 0 \) gives \( -s-2 < 0 \) (so  \( y_k^{\text{out}} \)= +1) or \( s \leq -2 \) gives \( -s-2 \geq 0 \) (so \( y_k^{\text{out}} \) = -1). 
   In both cases: \( -\operatorname{sign}(-s-2) = \operatorname{sign}(s) \).
\end{enumerate}

Equality to Raw BNN Bit:
\begin{equation}
(-1)^{{R_k}} \operatorname{sign}((-1)^{{R_k}} s - 2{R_k}) = \operatorname{sign}(s) \quad 
\end{equation}
\( \quad \quad \quad \quad \forall R_k \in \{0, 1\}, s \in 2\mathbb{Z}.\)

Thus (9) holds and combining (8):
\[
y_k^{\text{out}} = \operatorname{sign}(y_k - B_k) = (1)
\]

Therefore, for every input to the transformed BNN, the output matches the original BNN, establishing that the transformation and recovery process preserves the integrity of the output bit.

\section{ Protection Techniques and experiments \vspace{-2mm}}
\begin{figure}[!htbp]
 
\centerline{\includegraphics[width=0.4\textwidth]{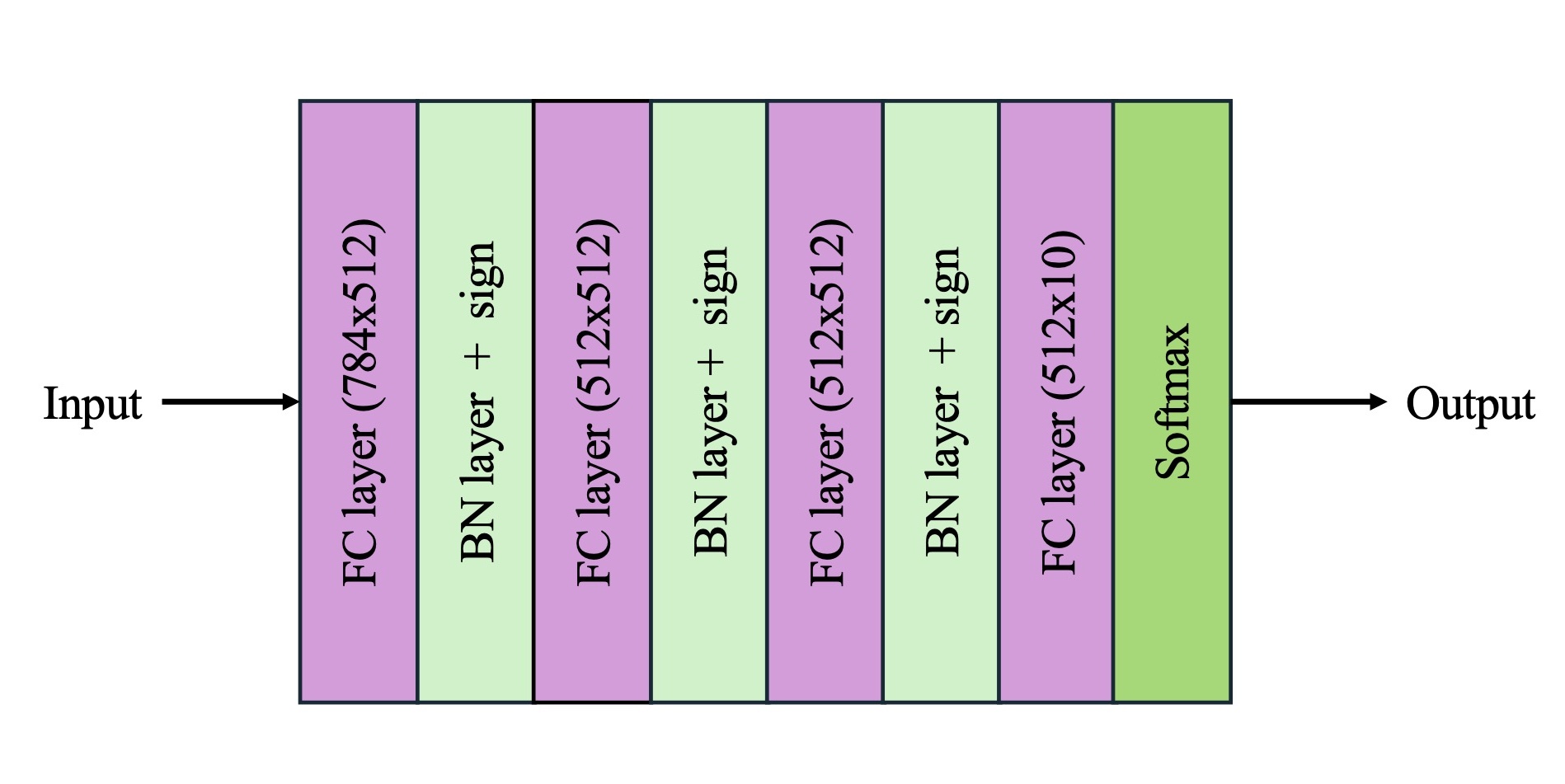}}
\caption{Architecture of the BNN trained on the MNIST dataset}
\label{bnn_mnist}
\vspace{-2mm}
\end{figure}

 Building on the formalization introduced in the previous section, we have categorized three distinct techniques for weight transformation: (i) inversion, (ii) swapping, and (iii) a combination of both techniques, wherein inversion and swapping are applied across either columns or rows. Our previous research \cite{rajendran2024puf,rajendran2024securing} has addressed the implementation of techniques (i) and (ii), while the current paper is dedicated to the exploration of technique (iii). To support our findings, we will employ the BNN topology detailed in Fig. \ref{bnn_mnist}, utilizing the MNIST dataset as a benchmark for validation. The responses generated by the PUF can be inadequate at times. It is advisable to expand them using cryptographic algorithms, such as key derivation functions, to achieve the desired length.

 \begin{figure}[!htbp]
\centerline{\includegraphics[width=0.45\textwidth]{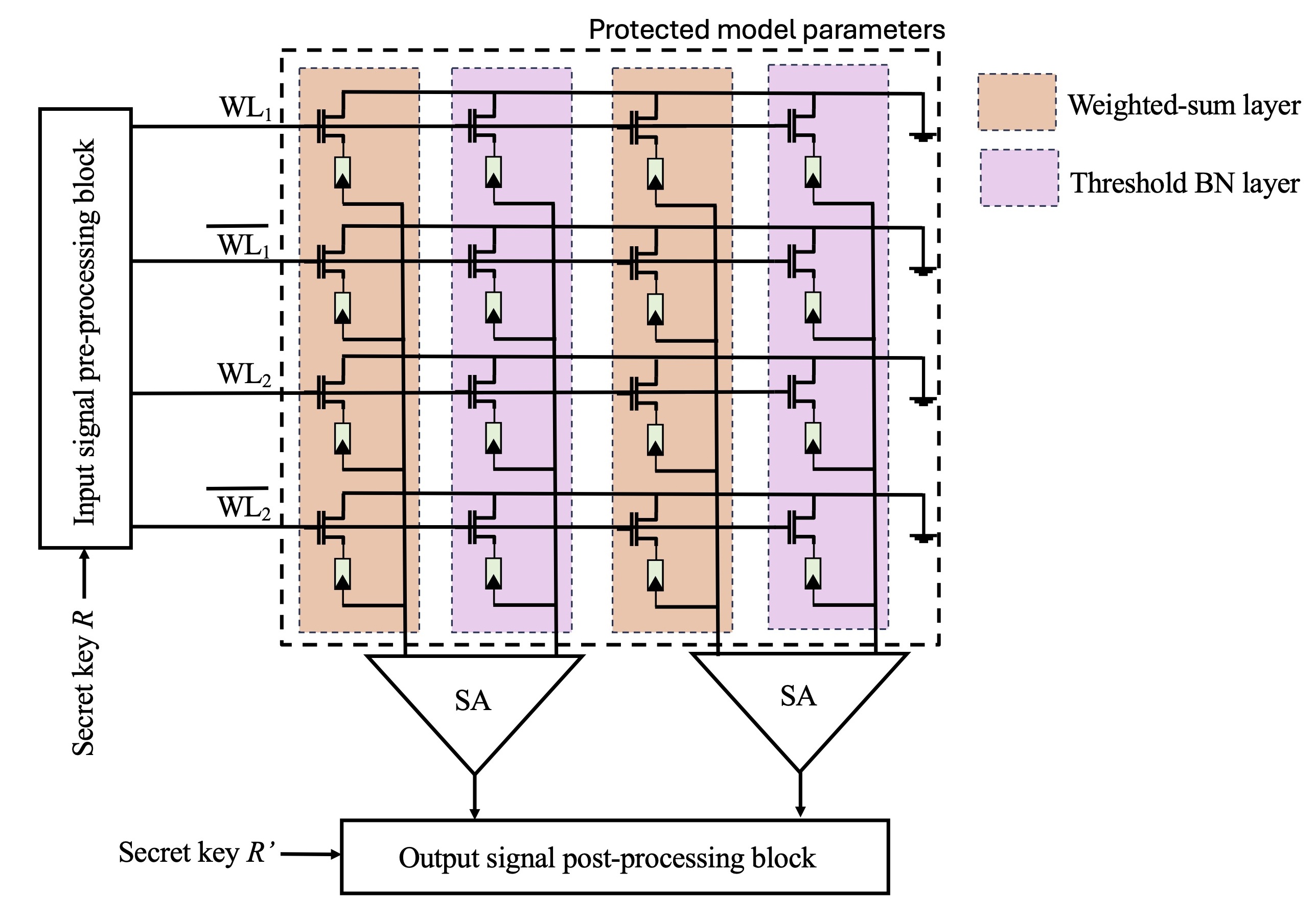}}
\caption{Implementation of protected BNN in RRAM crossbar}
\label{bnn_adc}
\vspace{-6mm}
\end{figure}

\subsection{Swap + Inversion along rows}

The original weights \( W \) of the BNN are transformed into protected weights \( W^* \) using a secret key \( R \), according to the equation \( W^* = P_R D_R W \). In this equation, \( P_R \) represents the row permutation matrix used for swapping, while \( D_R \) is the row diagonal matrix used for inversion. Both \( P_R \) and \( D_R \) are derived from the secret key \( R \). All permutation matrices discussed in this study are generated using the swap scheme outlined in our previous work \cite{rajendran2024securing}. The protected weights \( W^* \) are only mapped onto the crossbar. During inference, the input is also transformed using the secret key \( R \), resulting in \( X^* = P_R D_R X \), which is then applied to the crossbar. This process recovers the pre-activation values.

We require a secret key of length \( m \) for inversion along rows and a key of length \( \frac{m}{2} \) for swapping rows. Therefore, if we use separate keys for inversion and swapping, the total length of the secret key would be \( m + \frac{m}{2} \). Alternatively, we can utilize a single key of length \( m \), where the entire key can be used for inversion, and the first \( \frac{m}{2} \) can be reused for swapping.

We transformed the parameters of the BNN model according to this scheme and assessed the impact on accuracy without the secret key utilized for model protection. As illustrated in Table I, safeguarding both individual layers and all three layers leads to a reduction in accuracy when the secret key is not known.

\begin{table}[!htbp]
\caption{Inference accuracy with swapping and inversion along rows for MNIST BNN}
\centering
\begin{tabular}
{m{2cm}|m{2cm}|m{3cm}}
\hline
\hline
\textbf{Protected layers} &  \textbf{Secret key length (\( \frac{m}{2}  + m)\)} & \textbf{Inference accuracy without retrieving the Protected weights} \\
\hline
\hline
None & 0&96.74\% \\
\hline
FC 1 & 392 + 784 & 9.42\% \\
\hline
FC 2 & 256 + 512 & 9.85\% \\
\hline
FC 3 & 256 + 512 & 13.3\% \\
\hline
FC 1,2,3 & $392 + 2\times 256 + 784 + 2 \times 512 $  & 10.51\% \\
\hline
\hline
\end{tabular}
\label{table1}
\vspace{-4mm}
\end{table}

\subsection{Swap + Inversion along Columns}
To perform the transformation along the columns, we need to apply the transformation to both model parameters \( W \) and \( B \). For the parameter \( W \), the transformation is defined as \( W^* = W \cdot P_C \cdot D_C \), where \( P_C \) is the column permutation matrix and \( D_C \) is the column diagonal inversion matrix. Both \( P_C \) and \( D_C \) are derived from the secret key \( R \). For the parameter \( B \), the transformation is expressed as \( B^* = D_C (P_C B) + 2R \). The output is recovered after activation as \( P_C D_C y^{*(b)} \), where \( y^{*(b)} \) is the output derived from the operations on \( W^* \) and \( B^* \).

We need a secret key of length \( n \) for inversion and another key of length \( \frac{n}{2} \) for swapping columns. This implies that if we use distinct keys for each function, the total length of the secret keys would be \( n + \frac{n}{2} \). Alternatively, we could use a single key of length \( n \), utilizing the full length for inversion and reusing the first \( \frac{n}{2} \) for the swapping operation. This approach is similar to the one discussed in the previous case.

We also applied this scheme to the BNN model trained with the MNIST dataset, as detailed in Table II. We found that without the secret key used for model protection, the model's accuracy is significantly reduced, rendering it ineffective.

\begin{table}[!htbp]
\caption{Inference accuracy with swapping and inversion along columns for MNIST BNN}
\centering
\begin{tabular}
{m{2cm}|m{2cm}|m{3cm}}
\hline
\hline
\textbf{Protected layers} &  \textbf{Secret key length (\( \frac{n}{2} + n )\)} & \textbf{Inference accuracy without retrieving the Protected weights} \\
\hline
\hline
None & 0&96.74\% \\
\hline
FC 1 & 256+512 & 8.97\% \\
\hline
FC 2 & 256+512 & 7.66\% \\
\hline
FC 3 & 256+512 & 9.49\% \\
\hline
FC 1,2,3 & $3 \times (256+512)$ & 10.22\% \\
\hline
\hline
\end{tabular}
\label{table2}
\end{table}

\subsection{Row inversion + Column swap}
We can apply transformations along both rows and columns simultaneously. In this study, we will explore row inversion and column swapping as examples, though other combinations are also possible. To achieve this, we transform the model parameters as follows: \( W^* = D_R W P_C \) and \( B^* = P_C B \). During inference at runtime, we first transform the input using \( X^* = D_R X \) before applying it to the crossbar. Additionally, to recover the output after activation, we use \( P_C y^{*b} \).

To perform row inversion, we require a secret key of length \( m \), and for column swapping, we need a secret key of length \( \frac{n}{2} \). Therefore, the total length of the key needed for both operations is \( m + \frac{n}{2} \). Similar to the previous discussion, we also analyzed this scheme with a BNN model trained on the MNIST dataset. We found that without the secret key used for protection, the model's accuracy is very low, as detailed in Table III. The pre-processing and post-processing blocks consist only of MUX and XOR gates. Compared to the power consumption of the comparator, this is very small, adding less than 1\% to the total power, as discussed in our previous studies\cite{rajendran2024securing}.

\begin{table}[!htbp]
\vspace{-4mm}
\caption{Inference accuracy with row inversion and column swapping for MNIST BNN}
\centering
\begin{tabular}
{m{2cm}|m{2cm}|m{3cm}}
\hline
\hline
\textbf{Protected layers} &  \textbf{Secret key length (\( m + \frac{n}{2} \) }) & \textbf{Inference accuracy without retrieving the Protected weights} \\
\hline
\hline
None & 0&96.74\% \\
\hline
FC 1 & 784 + 256 & 10.8\% \\
\hline
FC 2 & 512 + 256 & 9.25\% \\
\hline
FC 3 & 512 +256 & 13.08\% \\
\hline
FC 1,2,3 & $784 + 2 \times 512 + 3 \times 256 $ & 9.67\% \\
\hline
\hline
\end{tabular}
\vspace{-4mm}

\label{table3}
\end{table}

\section{Conclusion and Future works}

We developed a method to protect any BNN implementation using secret keys, especially within in-memory computing architectures. We explored three techniques that can transform the model parameters using secret keys derived from PUFs to safeguard against theft attacks. Our findings indicate that without the secret key used for these transformations, the model's accuracy significantly reduces, falling below 15\%, which renders it ineffective. In future, it would be interesting to identify other activation functions and neural networks, beyond BNN, which can perform encrypted inference as well as training to achieve model privacy at lower cost compared to FHE.

\section{Acknowledgements}
This research is part of the IN-CYPHER programme supported by the National Research Foundation, Prime Minister’s Office, Singapore under its Campus for Research Excellence and Technological Enterprise (CREATE) programme.

\bibliographystyle{IEEEtran}

\bibliography{main}

\begin{thebibliography}{1}
\providecommand{\url}[1]{#1}
\csname url@samestyle\endcsname
\providecommand{\newblock}{\relax}
\providecommand{\bibinfo}[2]{#2}
\providecommand{\BIBentrySTDinterwordspacing}{\spaceskip=0pt\relax}
\providecommand{\BIBentryALTinterwordstretchfactor}{4}
\providecommand{\BIBentryALTinterwordspacing}{\spaceskip=\fontdimen2\font plus
\BIBentryALTinterwordstretchfactor\fontdimen3\font minus \fontdimen4\font\relax}
\providecommand{\BIBforeignlanguage}[2]{{%
\expandafter\ifx\csname l@#1\endcsname\relax
\typeout{** WARNING: IEEEtran.bst: No hyphenation pattern has been}%
\typeout{** loaded for the language `#1'. Using the pattern for}%
\typeout{** the default language instead.}%
\else
\language=\csname l@#1\endcsname
\fi
#2}}
\providecommand{\BIBdecl}{\relax}
\BIBdecl

\bibitem{kim2019memory}
H.~Kim~et al., ``In-memory batch-normalization for resistive memory based binary neural network hardware,'' in \emph{Proceedings of the 24th Asia and South Pacific Design Automation Conference}, 2019, pp. 645--650.

\bibitem{rajendran2024securing}
G.~Rajendran~et al., ``Securing binarized neural networks via puf-based key management in memristive crossbar arrays,'' \emph{IEEE Embedded Systems Letters}, vol.~17, no.~1, pp. 30--33, 2024.

\bibitem{galicia2023s3cure}
M.~E. Galicia~et al., ``" s3cure": Scramble, shuffle and shambles-secure deployment of weight matrices in memristor crossbar arrays,'' in \emph{Proceedings of the 2023 International Conference on Neuromorphic Systems}, 2023, pp. 1--8.

\bibitem{zou2022enhancing}
M.~Zou~et al., ``Enhancing security of memristor computing system through secure weight mapping,'' in \emph{2022 IEEE Computer Society Annual Symposium on VLSI (ISVLSI)}.\hskip 1em plus 0.5em minus 0.4em\relax IEEE, 2022, pp. 182--187.

\bibitem{wang2021low}
Y.~Wang~et al., ``A low cost weight obfuscation scheme for security enhancement of reram based neural network accelerators,'' in \emph{Proceedings of the 26th Asia and South Pacific Design Automation Conference}, 2021, pp. 499--504.

\bibitem{rajendran2024puf}
G.~Rajendran~et al., ``Puf-based lightweight authentication for binarized neural networks,'' in \emph{2024 IEEE Asia Pacific Conference on Circuits and Systems (APCCAS)}.\hskip 1em plus 0.5em minus 0.4em\relax IEEE, 2024, pp. 447--451.

\end{thebibliography}

\end{document}